\documentclass[review,5p,twocolumn,times]{elsarticle}

\usepackage{amssymb}
\usepackage{amsmath}

\usepackage{lipsum}
\usepackage{graphicx}
\usepackage{bm}
\usepackage{color}
\usepackage{times}
\usepackage{amsmath}
\usepackage{float}
\usepackage{hyperref}
\usepackage{ulem} 
\usepackage[dvipsnames]{xcolor}
\usepackage{array}

\newcommand{\beginsupplement}{%
	\setcounter{table}{0}
	\renewcommand{\thetable}{S\arabic{table}}%
	\setcounter{figure}{0}
	\renewcommand{\thefigure}{A\arabic{figure}}%
 	\setcounter{equation}{0}
	\renewcommand{\theequation}{S\arabic{equation}}%
	 \setcounter{page}{1}
	 \renewcommand{\thepage}{S\arabic{page}}%
	
}
\begin{document}

\begin{frontmatter}



\title{Hierarchical metamaterials with tunable flat bands, zero-frequency, and wavenumber gaps}

\author{Mohamed A. Elgamal}
\author{Osama R. Bilal}
\affiliation{School of Mechanical, Aerospace, and Manufacturing Engineering, University of Connecticut, Storrs , CT , 06269, USA.}

\begin{abstract}
Metamaterials are arrangement of basic building blocks that repeat in space, time, or both. These material systems serve as an excellent platform for controlling waves, such as engineering wavenumber band gaps, flat bands, and zero-frequency band gaps. However, combining one or more of these exotic features within the same unit cell design remains a challenge. Moreover, once a metamaterial is realized, its dispersive properties are usually fixed. In this work, we present a tunable passive hierarchical metamaterial capable of exhibiting wavenumber band gaps, flat bands, and zero-frequency band gaps within the same dispersion curve. Our metamaterial is composed of magnetic elements confined within a fixed magnetic boundary. The metamaterial can be tuned by adjusting the magnetic boundary, which in turn can alter the lattice periodicity. We open wavenumber band gaps by incorporating magnetic coupling within the unit cell elements, resulting in negative physical stiffness. The tunability of the magnetic coupling also enables complete flattening of the dispersion bands. Moreover, the ground stiffness within our unit-cell design causes the opening of zero-frequency band gaps. We present our approach through a combination of analytical, numerical, and experimental methods. The analytical framework provides a blueprint for obtaining each of these exotic dispersion characteristics. The numerical analysis, using both linear and nonlinear models, validates our analytical predictions, which we further confirm through experimental demonstrations. Our work opens the door to exploring magnetic tunability and hierarchy in engineering metamaterial systems with exotic properties that can be harnessed in advanced acoustic and mechanical devices.
\end{abstract}

%


\end{frontmatter}       



\section{\label{sect:Intro} Introduction}

The dispersion relation, or the correlation between wavenumber and frequency, can reveal multiple wave transmission properties of materials and structures. The tailoring of dispersion characteristics, such as the shape of the dispersion curves and their frequency ranges, remains a challenge in wave engineering. Metamaterials are among the most promising platforms for dispersion engineering as they exhibit exotic wave behavior such as lensing, focusing, bending, or even cloaking \cite{kadic20193d}. Metamaterials are engineered materials with unique patterns that repeat in space, time, or both. These patterns can be designed to influence the spacing between dispersion branches, giving rise to band gaps or frequency ranges where waves are not allowed to propagate. In addition to opening band gaps at finite frequency values, the unit cell design can affect the slope of the dispersion branches. The slope of the dispersion curve dictates the group velocity of the propagating wave at a given frequency. Metamaterials can enable the engineering of highly localized waves with zero group velocities for all wave numbers at a given frequency (i.e.,  flat dispersion band).  Furthermore, the unit cell design can influence the lowest value of the first dispersion branch both in frequency and wavenumber. Resulting in exotic dynamical characteristics such as wavenumber band gaps or zero frequency band gaps. 

There exists an extensive body of literature on tailoring the dispersion curves in engineered materials for a specific objective. For example,   designing a unit cell with flat dispersion branches was originally introduced in electronic systems \cite{tamura2002flat,sutherland1986localization,misumi2017new,hase2018possibility}, and was extended to optics \cite{yang2024realization,leykam2018artificial,leykam2013flat,apaja2010flat,mukherjee2015observation,shen2010single,nguyen2018symmetry}, acoustics \cite{riva2025creating,zheng2014acoustic,dubois2019acoustic,wu2016flat,karki2023non,zhang2025sound,shen2022observing,ye2024experimental,han2025all}, and elastic waves \cite{wang2013lamb,riva2025non,samak2024direct,qian2020observation,li2023seismic} as means of controlling light, sound, and vibrations. These flat bands have many potential applications in optics. For example,  lasing \cite{longhi2019photonic}, slow light propagation \cite{li2008systematic}, and mode conversion \cite{kim2022mode}. Acoustic flat bands have also been proposed for sound lensing \cite{dubois2019acoustic}, sonar cloaking \cite{zheng2014acoustic}, ultrasensitive measurements \cite{eldegail2025design}, energy harvesting \cite{park2024all,deng2024dual}, and selective mode localization \cite{shen2022observing}. Another exotic wave property that can be encoded into the dispersion curves is zero frequency band gaps. Unlike conventional band gaps, zero-frequency band gap materials have all dispersion branches starting from a finite frequency value that is greater than zero. A zero frequency band gap can be opened by introducing ground stiffness to the unit cell \cite{ren2020active,yu2012propagation,yang2024experimental} in the form of clamped barriers \cite{luo2025zero}, fixed magnetic boundary \cite{stenseng2025bi,eichelberg2022metamaterials,watkins2022harnessing}, coupling between spinning motion and longitudinal motion \cite{oh2018zero}, optomechanical interactions \cite{lemkalli2025space}, or  negative stiffness and moment of inertia \cite{yang2021ultrawide}. Metamaterials with zero frequency band gaps can have applications in sensing and attenuating ultra-low frequency such as seismic waves \cite{watkins2020demultiplexing, huang2025seismic,wang2021periodic,wang2025broadband}. An additional exotic feature that a metamaterial's dispersion can exhibit is a wavenumber band gap. It is important to distinguish wavenumber band gaps from conventional frequency band gaps. A band gap within the frequency spectrum is a range of frequencies at which the an excited wave attenuates exponentially, regardless of the wavenumber. In contrast, a wavenumber band gap refers to a range of wavenumbers for which wave propagation is inherently unstable, regardless of the frequency. A wavenumber band gap in metamaterials can be engineered using several techniques, including excessive damping \cite{hussein2010band}, temporal modulation \cite{chong2024modulation,kim2023dynamics}, Wills-type lattices \cite{al2025onset}, and  negative physical stiffness \cite{samak2025observation}. 

An intriguing approach to designing metamaterial unit cells is through hierarchy, where the system is composed of structural elements that themselves possess internal structures \cite{gatt2015hierarchical}. Hierarchy can augment various mechanical properties of metamaterials (e.g.,  stiffness, strength, and energy absorption) through the incorporation of multiple structural scales \cite{ qin2025hierarchical, mizzi2020lightweight, song2024design, zhang2022dual,zhu2025bio,pan2025hierarchical,li2024hierarchical}. While hierarchy can enhance functionality, once the metamaterials are fabricated, their properties are usually fixed. A growing trend in the literature considers the tunability of metamaterials through the utilization of various external stimuli such as magnetic fields \cite{perez2025reprogrammable,sim2025selective,deng2024locally, allein2016tunable, robillard2009tunable, fan2024tunable,wang2025tunable}, electrical fields \cite{cha2018electrical}, thermal tuning  \cite{zhang2025re, liu2024tunable,zha2025negative,chuang2019bandgap}, tuning with light \cite{meng2025light,gliozzi2020tunable}, mechanical loads \cite{zhang20253d,zhang2021hierarchical,dudek2022micro,hua2024design,berwind2018hierarchical,li2021programmable,surjadi2019mechanical}, gear-based metamaterial \cite{wang2025double}, unit cell mass tuning \cite{zhang2024tunable}, piezoelectricity \cite{shao2022tunable, ren2020active, sepehri2022active},  pneumatic actuation \cite{liu2025mechanical,ochoa2025reconfigurable}, and ferroelectric-based metamaterial \cite{roshdy2024observation,li2021bulk, roshdy2023tunable}. However, a metamaterial design that can combine all these dynamical exotic effects while being tunable remains elusive. In this paper, we utilize a tunable hierarchical design approach for metamaterials to sculpt their dispersion curves into (1) wavenumber band gaps, (2) flat bands, and (3) zero-frequency band gaps. Through a combination of analytical modeling, numerical simulations, and experimental validation, we report the first observation of such exotic phenomena in a tuned passive hierarchical system.

\begin{figure*}[!h] 
\centering\includegraphics[scale = 1.01]{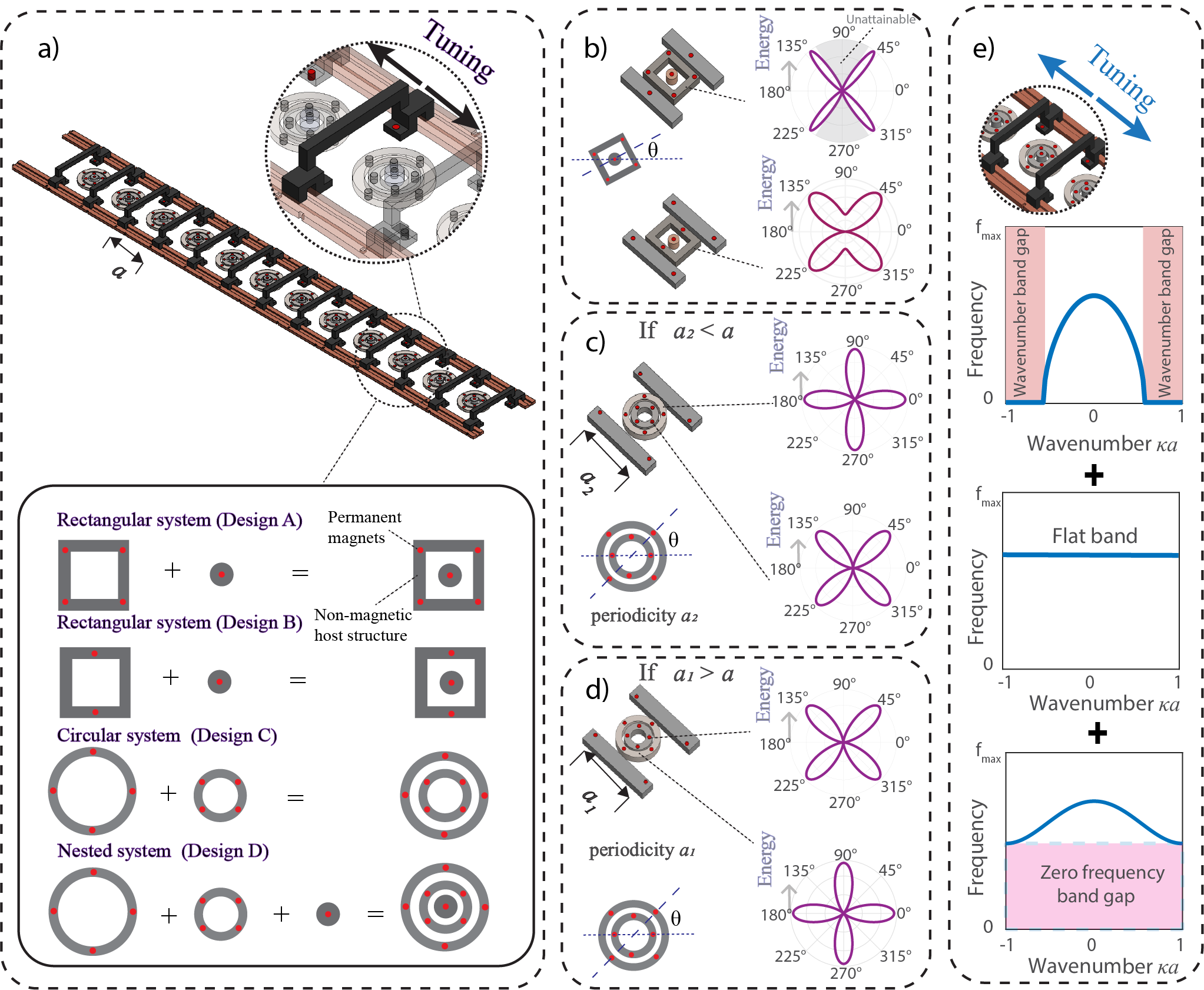}
\caption{\label{fig:Concept}\textbf{Hierarchical metamaterial unit cell designs, orientation, and dispersive properties.} (a) Schematic of the tunable metamaterial designs (rectangular and circular frames with inner resonators). (b) The minimum energy positions of both the rectangular frames (Design A and Design B). The minimum energy position for the circular moving frame and its inner resonator (Design C and Design D) in the case of (c) small lattice constant. (d) large lattice constant. (e) By varying the periodicity of the lattice, we can achieve dispersion curves with wavenumber band gap, flat bands, and zero frequency band gap.}
\end{figure*}
\section{\label{sect:Intro} SYSTEM DESIGN AND CONCEPTUAL FRAMEWORK}

Our unit cells are constructed of nested resonators with tunable periodicity. Each resonator is surrounded by a magnetic boundary that can be tuned to change the lattice periodicity. Each resonator is composed of embedded permanent magnets within a non-magnetic host structure. All the magnets within both the boundary and the hierarchical elements are aligned in the same polarity direction (i.e., all north-poles facing upward). As all permanent magnets exist in the same plane, they act as monopoles in repulsion. We engineer the level of hierarchy within the unit cell from a simple square frame surrounding a circular disk, to a more sophisticated design with three concentric rings (Fig. \ref{fig:Concept}a). To open zero-frequency band gaps, we utilize the fixed magnetic boundary which acts as ground stiffness. To induce wavenumber band gaps, we tune the periodicity such that the inter-stiffness is sufficiently negative. To flatten the dispersion bands, we increase the coupling stiffness between the resonators while increasing the lattice periodicity (See Appendix for more details).

Figure \ref{fig:Concept}(a) includes our tunable metamaterial design with four different levels of complexity.
\textbf{Design A} a unit cell composed of a rectangular frame (30 mm x 25 mm) with four embedded cylindrical magnets at its corners (3 mm in diameter with 3 mm in height) and a central inner disc (8 mm in diameter and 3 mm in height) acting as a resonator.
\textbf{Design B} a rectangular frame (30 mm x 30 mm) with two embedded cylindrical magnets at the middle of its length and a central inner disc (8 mm in diameter and 3 mm in height) acting as a resonator.
\textbf{Design C } a circular frame (30 mm in diameter) with four embedded magnets spaced at equidistant points along the perimeter. In contrast to designs A and B, here the inner resonator is a ring-shaped structure (18 mm in diameter) hosting four embedded magnets rather than one.
\textbf{Finally, Design D } is a nested system composed of three circular elements: an outer ring (30 mm in diameter) with four equally-spaced  embedded magnets, a middle ring (18 mm in diameter) hosting an additional four magnets, and an inner disc (8 mm in diameter) with a single embedded magnet. All the embedded magnets within our design are 3x3 mm cylindrical magnets except for the inner disk within designs B and D, which is 1 mm in height.

Each of the four designs is floated on an air-bearing table to minimize friction. As a result, the elements within each design are free to rotate and self-align based solely on the magnetic repulsion forces. To determine the orientation corresponding to the minimum energy configuration of the masses within the system, we employ the following energy model \cite{watkins2021exploiting, samak2024evidence}:

\begin{align}
    E = - \int_{0}^{\theta} \sum_{s=1}^{N} \Bigg( 
    \sum_{l=1}^{f} F_{s,L}^{y} \cdot r \cdot \cos\left(\theta + \frac{2\pi s}{N}\right) 
    \nonumber \\
    - \sum_{l=1}^{f} F_{s,L}^{x} \cdot r \cdot \sin\left(\theta + \frac{2\pi s}{N}\right) 
    \Bigg) \, d\theta,
\end{align}

where $N$ is the total number of magnets in a given mass, $s$ is the magnet index within a unit cell $n$, $l$ is the index of the interacting magnets in neighboring cells $n+1$ and $n-1$, the internal resonator of the unit cell $n$, and the magnets at the boundary. Therefore, $f$ represents the total number of magnets interacting with each magnet inside the mass of interest (i.e., $f=11$ in case of the rectangular frame and $f=16$ in case of the ring).  $r$ is the distance between the center of the magnet and the center of the considered mass of interest. Finally, $ F_{s,l}^y $ and $ F_{s,l}^x $ are the total forces acting on each magnet in the $ y $- and $ x $-directions, respectively. In the case of the rectangular outer mass, the lowest energy state is at $0^{\circ}$ orientation for the outer frame (Fig.~\ref{fig:Concept}(b), while the orientation angle is irrelevant for the inner disk.   In the case of the ring, the minimum energy position is a function of the lattice periodicity. For small lattice constants (below a critical $a$ value), the minimum energy orientation is  45\textdegree~ of the outer ring and 0\textdegree~ for the inner ring. As periodicity increases beyond $a$, the minimum energy state shifts from 45\textdegree~  to 0\textdegree~. Consequently, the orientation of the inner resonator also changes from 0\textdegree~  to 45\textdegree~. In other words, the outer ring and the inner ring rotate relative to each other to minimize their energy in a complementary way, resembling the motion of a gear system.

To tune the dynamical behavior of our metamaterials, we design the lattice such that its periodicity can be easily modified post-fabrication. Tunable periodicity is crucial as it directly affects the dispersive properties of the metamaterials (e.g., the width and the central frequency of the band gaps). We tune the periodicity by incorporating adjustable arms with embedded magnets that can slide along the fixed outer frame. By repositioning the arms, and consequently the magnets within, the spacing between the unit cells changes without the need for a complete redesign of the unit cells. Through the precise control of the lattice's periodicity, we realize  wavenumber band gaps, flat bands, and zero-frequency band gaps (Fig.\ref{fig:Concept}(e)).

\section{\label{sect: Intro} Dynamic Modeling and Governing Equations}

\begin{figure} [!b]
\centering
\includegraphics{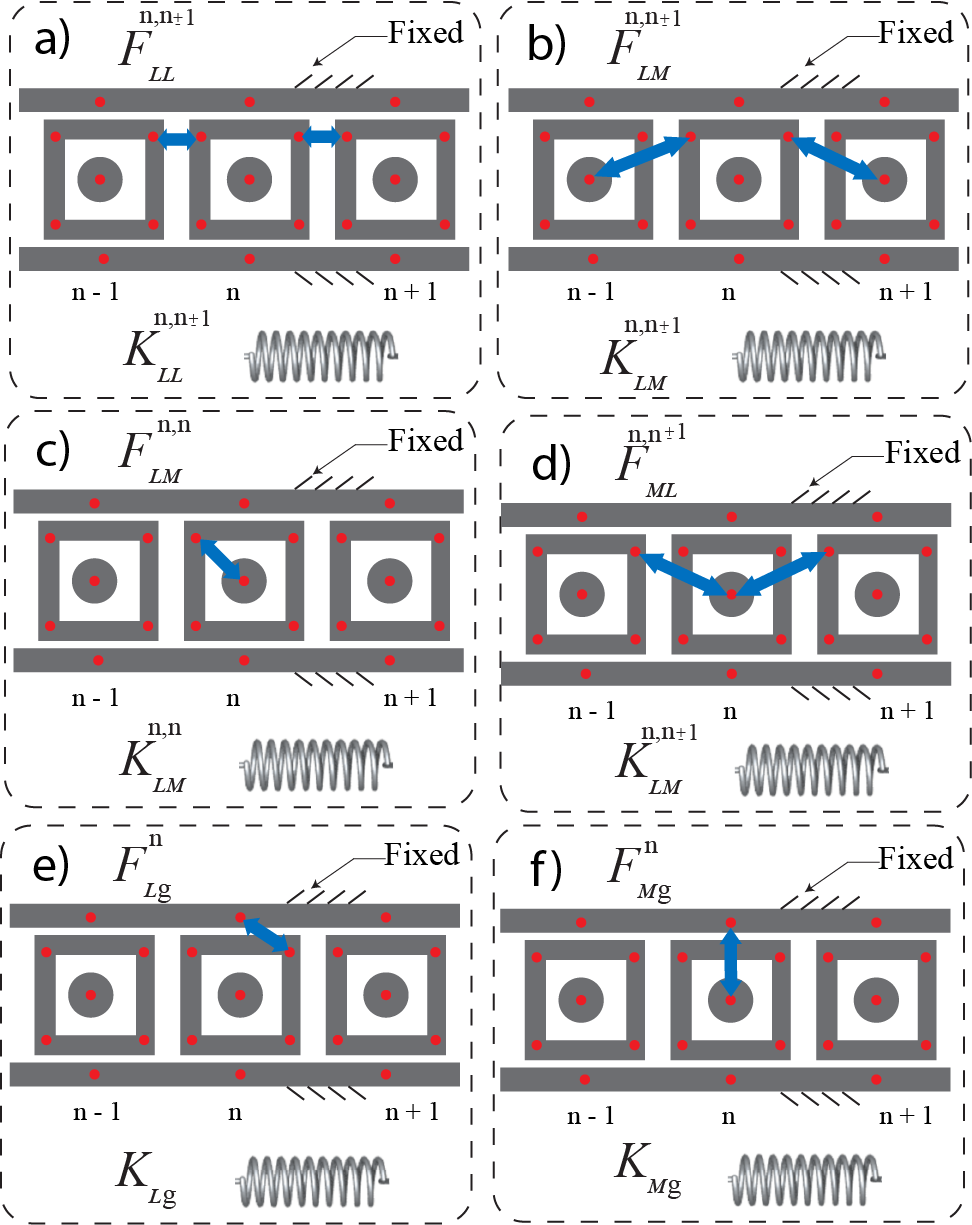}
\caption{\label{Interactions}\textbf{Magnetic forces between the moving rectangular masses, resonators, and the fixed outer-frame: nonlinear forces and linear approximations.} The forces between  (a) the rectangular moving frame in the unit cell and its neighboring unit cells,  (b) the rectangular moving frame and the resonators in the neighboring unit cells, (c) the rectangular moving frame and the resonator in the same unit cell, (d) the resonator and the rectangular moving frame in the neighboring unit cells, (e) the rectangular moving frame and the fixed outer-frame (i.e. ground forces), and  (f) the resonator and the the fixed outer-frame. }
\end{figure}

To understand the dynamics of our metamaterials, we consider both the nonlinear equations that govern their dynamics in addition to a linearized version of the model. We utilize the complete nonlinear model to capture the dispersive properties of the system that cannot be simulated by our linear analysis alone, in particular the dynamics of the finite system in the presence of wavenumber band gaps. Each mass can move in two directions (i.e., shear and longitudinal) resulting in two degrees of freedom per mass. For designs A, B and C hierarchical unit cells, we have 4 degrees of freedom (DoF), while for design D hierarchy we have 6 DoF. We consider all possible magnetic interactions between the magnets embedded in the main mass, the resonator, and the fixed boundary. Figure ~\ref{Interactions} illustrates the magnetic couplings between the magnets located in the main mass, the resonator, and the boundary in design A. We begin with the full nonlinear equations as \cite{chong2021nonlinear,chong2024modulation}:\\

\noindent \textbf{For the main mass:}
 \begin{equation}    
  m_1 \ddot{u_j}^n +C \dot{u_j}^n = F_{LL}^{n,n\pm1} + F_{LM}^{n,n} + F_{LM}^{n,n\pm1} + F_{Lg}^n + F_{ex}^n
\end{equation}

\noindent \textbf{For the resonator:}
 \begin{equation}
    m_2 \ddot{v_j}^n+C \dot{v_j}^n = F_{ML}^{n,n} + F_{ML}^{n,n\pm1} + F_{Mg}^n 
\end{equation}

where  $m_1$  is the mass of the outer moving frame and $ m_2 $ is the mass of the resonator, $ u^{n}_{j} $ is the in-plane motion for mass $m_1$ in unit cell $n$; $ j = 1 $ represents longitudinal motion and $ j = 2 $ represents shear motion. $ v^{n}_{j} $ is the in-plane motion for mass $m_2$ in unit cell $n$. The term $ F_{LL}^{n,n\pm1} $ denotes the repulsive force between the main mass ($m_1$) in unit cell $ n $ and the main masses in neighboring unit cells $ n+1 $ and $ n-1 $ (Fig. \ref{Interactions}a). $F_{LM}^{n,n\pm1} $ denotes the repulsive force between the main mass in unit cell $ n $ and the resonators in the neighboring cells $ n+1 $ and $ n-1 $ (Fig. \ref{Interactions}b). $ F_{LM}^{n,n} $ denotes the repulsive force between the main mass in unit cell $ n $ and the resonator in the same unit cell (Fig. \ref{Interactions}c).  $ F_{ML}^{n,n\pm1} $ denotes the repulsive force between the resonator in unit cell $ n $ and the main masses in the neighboring unit cells (Fig. \ref{Interactions}d). $ F_{Lg} $ and $ F_{Mg} $ represent the repulsive forces between the main mass and the fixed frame, and the resonator and the fixed frame, respectively (Fig. \ref{Interactions}e and f). 
The magnetic repulsive force between any two magnets in the system is modeled as  a power-law in the form:
\begin{equation}
    F = A \cdot d_{1 \rightarrow 2}^\gamma
\end{equation}

where $F$ is the magnitude of the repulsive force, $d$ is the center-to-center distance between the two magnets, $A$ and $\gamma$ are experimentally determined constants that depend on the magnet type, size, and configuration. As the magnets are free to move in-plane, the distance $d$ between any two magnets varies with time and depends on their instantaneous positions. If we denote the planar positions of any two magnets by $(x_1, y_1)$ and $(x_2, y_2)$, then the instantaneous distance between them is given by:
\begin{equation}
    d_{1 \rightarrow 2}(t) = \sqrt{(x_2(t) - x_1(t))^2 + (y_2(t) - y_1(t))^2}
    \label{eq:distance}
\end{equation}

We use Eq.\ref{eq:distance} to evaluate the time-dependent magnetic repulsive force between the system components (i.e., main masses, resonators, and fixed boundary).

In addition to the full nonlinear equations of motion, we simplify our model using equivalent linear spring forces as an approximation for the magnetic forces between the interacting magnets. Specifically, the inter-magnet stiffness between two moving magnets can be expressed as \cite {samak2024direct}:
\begin{equation}
\mathbf{k}_{\text{in}} = - \sum_{s=1}^{n_a} \left( f_{,d} \, \mathbf{e}_s \otimes \mathbf{e}_s + \frac{f(d_s)}{d_s} \left( \mathbf{I} - \mathbf{e}_s \otimes \mathbf{e}_s \right) \right)
\end{equation}
where $n_a$ is the number of magnets in the moving mass of interest (i.e., neighbor mass or resonator), $\mathbf{e}_s$   denotes the unit direction vector, $d_s$ is the distance of the $s^\text{th}$ magnet pair, $f(d)$ is the magnetic force as a function of distance $d$, $f_{,d}$ is the first derivative of the repulsive force, and $\mathbf{I}$ is the identity matrix. The operator $\otimes$ represents the dyadic (outer) product.
 Similarly, the stiffness contribution from boundary (i.e., fixed) magnets can be defined as \cite{samak2024direct}:
\begin{equation}
\mathbf{k}_{\text{g}} = - \sum_{s=1}^{n_b} \left( f_{,d} \, \mathbf{e}_s \otimes \mathbf{e}_s + \frac{f(d_s)}{d_s} \left( \mathbf{I} - \mathbf{e}_s \otimes \mathbf{e}_s \right) \right)
\end{equation}
where $n_b$ is the number of boundary magnets and $\mathbf{e}_s$ and $d_s$ are the unit direction vector and distance for the $s^\text{th}$ magnet-pair, respectively. Therefore, we can set our linear equations to the following formulation:
\newline
\newline
\textbf{For the main mass:}
\begin{align}
m_1 \ddot{u}_j^i 
&+ K_{LL}^{n,n+1} (u_j^n - u_j^{n+1}) 
+ K_{LL}^{n,n-1} (u_j^n - u_j^{n-1}) \nonumber \\
&+ K_{LM}^{i,i} (u_j^n - v_j^n) 
+ K_{LM}^{i,i+1} (u_j^n - v_j^{n+1}) \nonumber \\
&+  K_{LM}^{i,i-1} (u_j^n - v_j^{n-1}) \nonumber \\
&+ K_{Lg} u_j^n 
= F_{\text{ex}}^n
\label{eq: Linear-main mass}
\end{align}    
\newline
\textbf{For the internal resonator:}
\begin{align}
m_2 \ddot{v}_j^n 
&+ K_{LM}^{n,n} (v_j^n - u_j^{n}) 
+ K_{LM}^{n,n+1} (v_j^n - u_j^{n+1}) \nonumber \\
&+ K_{LM}^{n,n-1} (v_j^n - u_j^{n-1}) 
+ K_{Mg} v_j^n 
= 0
\label{eq: Linear-resonator}
\end{align}    
After setting the linear equations of motion, we can obtain the analytical solution for the dispersive properties of the system by applying Bloch's theorem. Assuming harmonic wave solutions in the form:

\begin{equation}
u_j^n = U_j e^{i(\omega t - \kappa n a)}, \qquad v_j^n = V_j e^{i(\omega t - \kappa n a)},
\label{eq: Bloch solution}
\end{equation}

where $ U_j $ and $ V_j $ are the displacement amplitudes, $ \omega $ is the angular frequency, $ \kappa $ is the wavenumber, and $ a $ is the lattice constant. Substituting equations \ref{eq: Bloch solution} into the linear equations of motion \ref{eq: Linear-main mass}, \ref{eq: Linear-resonator} leads to the following eigenvalue problem:

\begin{equation}
[-\omega^2 \mathbf{M} + \mathbf{K}(\kappa)] \mathbf{A} = \mathbf{0},
\end{equation}

where $ \mathbf{A} = \begin{bmatrix} U_j \\ V_j \end{bmatrix} $, $ \mathbf{M} $ is the mass matrix, and $ \mathbf{K}(\kappa) $ is the stiffness matrix, which depends on the wavenumber $ \kappa $ due to the periodic nature of the structure. Solving the eigenvalue problem yields the dispersion relation $ \omega(k) $, which characterizes the wave propagation behavior in the system.

\section{\label{sect: Intro} Results and Discussion}
\begin{figure}[!h]
\centering
\includegraphics{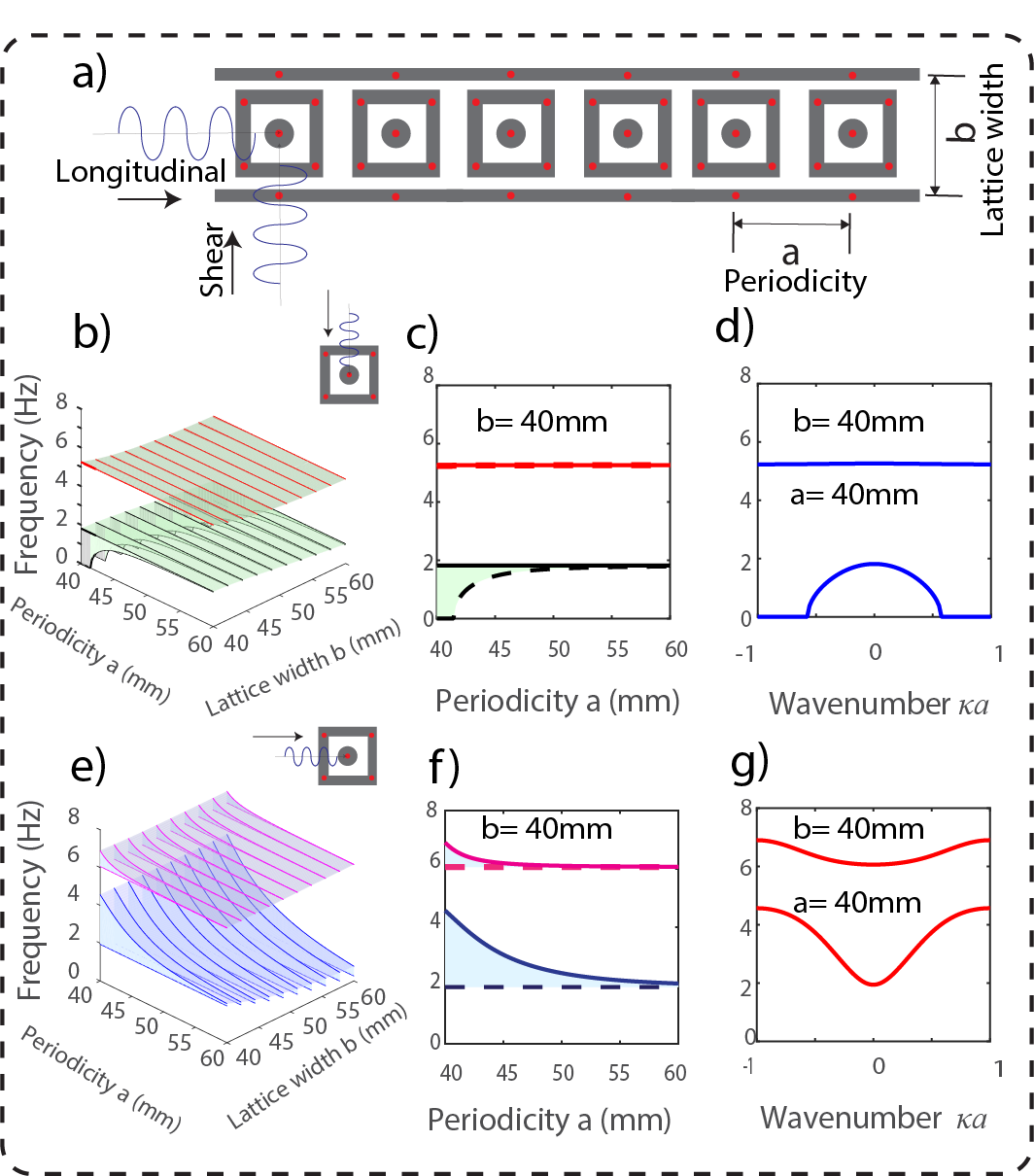}
\caption{\label{Parametric rectangular}\textbf{A study of the rectangular system's parameters in Design A} (a) A schematic of the rectangular system indicating the directions of the longitudinal and shear excitations. Three-dimensional parametric sweep of (b) shear modes and (e) longitudinal modes, showing the effect of varying both the lattice width and periodicity on the frequency range of transmission.  Dispersion frequency band width as a function of lattice periodicity with unit cell width = 40 $\mathrm{mm}$ for (c) shear modes and (f) longitudinal modes.   
Dispersion relation (frequency vs. wavenumber) for unit cell width and length $ a = b = 40 \, \mathrm{mm} $  for (d) shear modes and (g) longitudinal modes. }
\end{figure}

To explore the design search space for our metamaterials, we calculate the dispersion curves with different unit cell parameters for each of the proposed hierarchical designs. To achieve our objective of inducing a wavenumber band gap, a flat band, and a zero-frequency band gap, it is essential to tailor the physical stiffness of the system. We vary the key geometric parameters of the system that affect the distance between the different magnets in the unit cell: the \textit{periodicity,}  $a$, and the \textit{lattice width,} $ b $. We start with the rectangular unit cell design.  For a given unit cell, each dispersion plot includes two dispersion branches for the longitudinal direction and two branches for the shear direction. To better visualize the results, we separate the shear branches (Fig. \ref{Parametric rectangular}b-d) from the longitudinal ones (Fig. \ref{Parametric rectangular}e-g) in two different dispersion plots.

We record the evolution of the shear dispersion branches as a function of $a$ and $b$ (Fig. \ref{Parametric rectangular}b). Notably, by \textit{reducing the periodicity},  $a$, below 41 mm for a unit cell width of 40 mm, we achieve sufficient \textit{negative} inter-stiffness $ \mathbf{k}_{\text{in}} $, causing the emergence of a wavenumber band gap at the lowest dispersion branch. By increasing the lattice width, $b$, the critical value of $a$ at which wavenumber band gaps start to open increases to 46 mm at b = 60 mm. For the magnets to have physical negative stiffness, we consider the different forces and their directions within the meta-structure. When the magnetic disk moves in the shear direction, the repulsive forces exerted by the neighboring cells on the disk are in the same direction as the displacement. For a conventional mass-spring system, the restoring spring-force acts in the opposite direction of the motion. This unconventional behavior, where the force reinforces rather than resists the displacement, is effectively equivalent to a physical spring with negative stiffness \cite{samak2025observation}. Figure \ref{Parametric rectangular}(c) shows a cross section of the 3D parametric sweep for a lattice with fixed width, $b$ = 40 mm. As the periodicity increases, the dispersiveness of the lowest transmission band decreases. In other words, the dispersion branch becomes flatter due to the reduction in the inter-stiffness between each cell and its neighbors. The upper shear branch is flat for all the considered $a$ values. The parametric study also reveals that for a given value of $a$ = $b$ = 40 mm, we can obtain a simultaneous wavenumber band gap and a flat band within the same dispersion curve (Fig.~\ref{Parametric rectangular}d).

We also keep track of the evolution of the longitudinal branches as a function of the lattice parameters $a$ and $b$ (Fig.  \ref{Parametric rectangular}e), in a similar way to that of the shear branches. For both longitudinal branches, as the periodicity increases, the dispersion branches lose their dispersive properties and become flat (Fig. \ref{Parametric rectangular}f). We note that both longitudinal and shear motions have an upper flat band, albeit at different frequencies. An example dispersion curve for $a$ = $b$ = 40 mm shows two dispersive bands for longitudinal waves with a clear zero-frequency band gap (Fig. \ref{Parametric rectangular}g). It is important to note that for $a$ = $b$ = 40 mm, we have a single dispersion curve with (1) a wavenumber band gap for the lowest shear dispersion branch, (2) a flat band for the second shear branch, and (3) a zero-frequency band gap for all longitudinal modes.

\begin{figure*}
\centering
    \includegraphics[scale = 0.98]{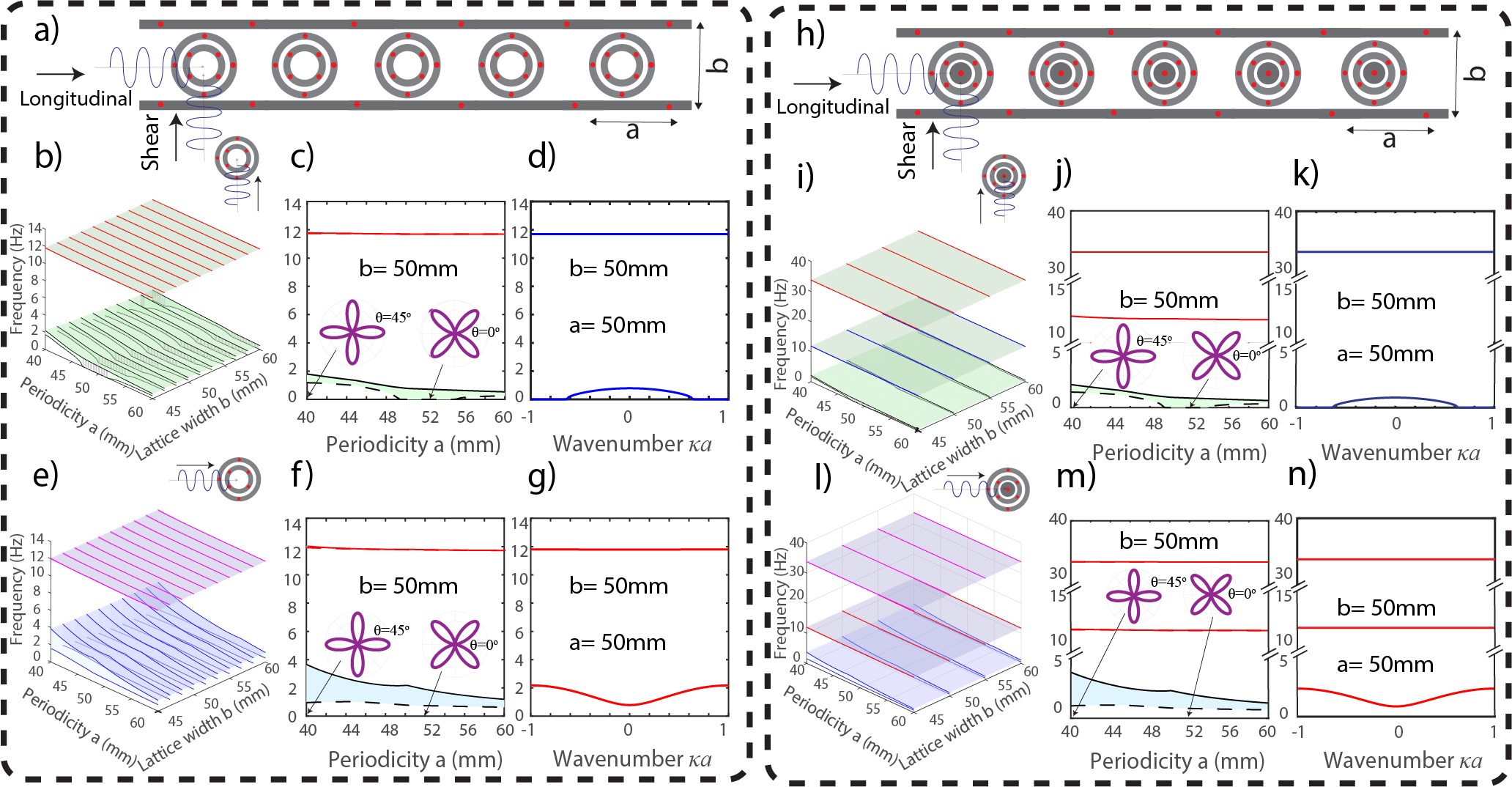} 
\caption{\label{Parametric circular}
\textbf{A study of the rectangular system's parameters in Designs C and D} A schematic of the circular system with (a) double circular design and (h) triple circular nested design. Three-dimensional parametric sweep of (b\& e) Design C and (i\& l) Design D in both shear and longitudinal modes, respectively, showing the effect of varying both the lattice width and periodicity on the frequency range of transmission.  Dispersion frequency band width as a function of lattice periodicity with unit cell width = 50 $\mathrm{mm}$ for (c\& f) Design C and (j\& m) Design D for shear and longitudinal modes, respectively.   
Dispersion relation (frequency vs. wavenumber) for unit cell width and length $ a = b = 50 \, \mathrm{mm} $  for (d\& g) Design C and (k\& n) Design D for shear and longitudinal modes, respectively.}
\end{figure*}

\begin{figure*}[!h]
    \centering
    \includegraphics[scale=2.28]{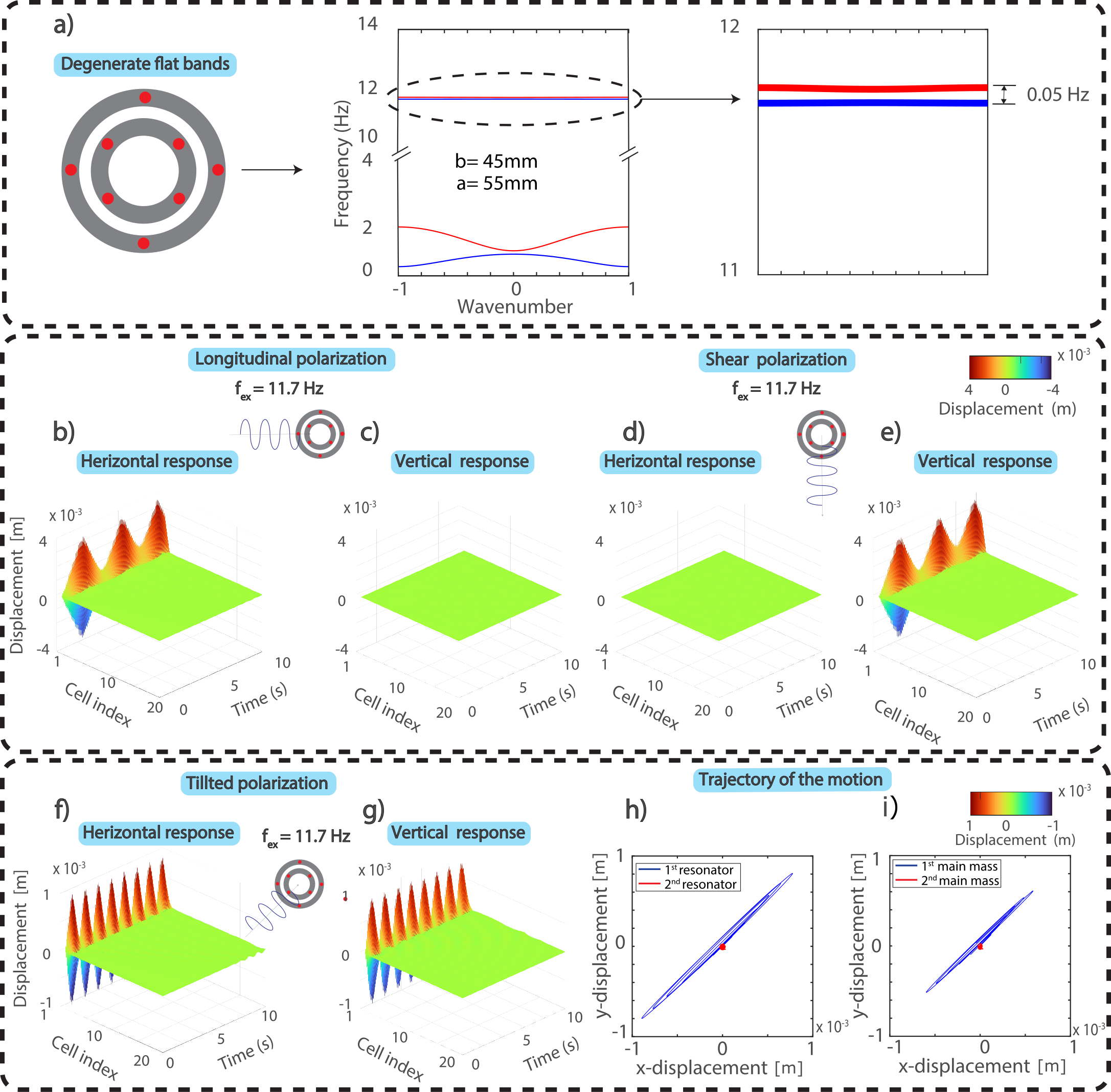}
    \caption{\label{degenerate flat band}\textbf{Degenerate flat band and polarization independent localization}. (a) Unit-cell schematic (left), dispersion diagram (center) showing two degenerate flat bands associated with the inner resonator at $f = 11.7$~Hz, and a zoomed view (right) showing $\Delta f = 0.05$~Hz. Time responses of the inner resonators under longitudinal polarization (b) in the $x$ direction  and (c) in the $y$-direction. Time responses of the inner resonators under shear polarization  (d) in the $x$ direction  and (e) in the $y$-direction. Time responses of the inner resonators under tilted (45$^\circ$) polarization (f) in the $x$- direction and (g) in the $y$-direction. Displacement trajectories of (h) the first and second inner resonators and (l) the first and second main masses under tilted polarization.}
\end{figure*}

\begin{figure*}[!h]
\centering
\includegraphics[width=0.98\linewidth]{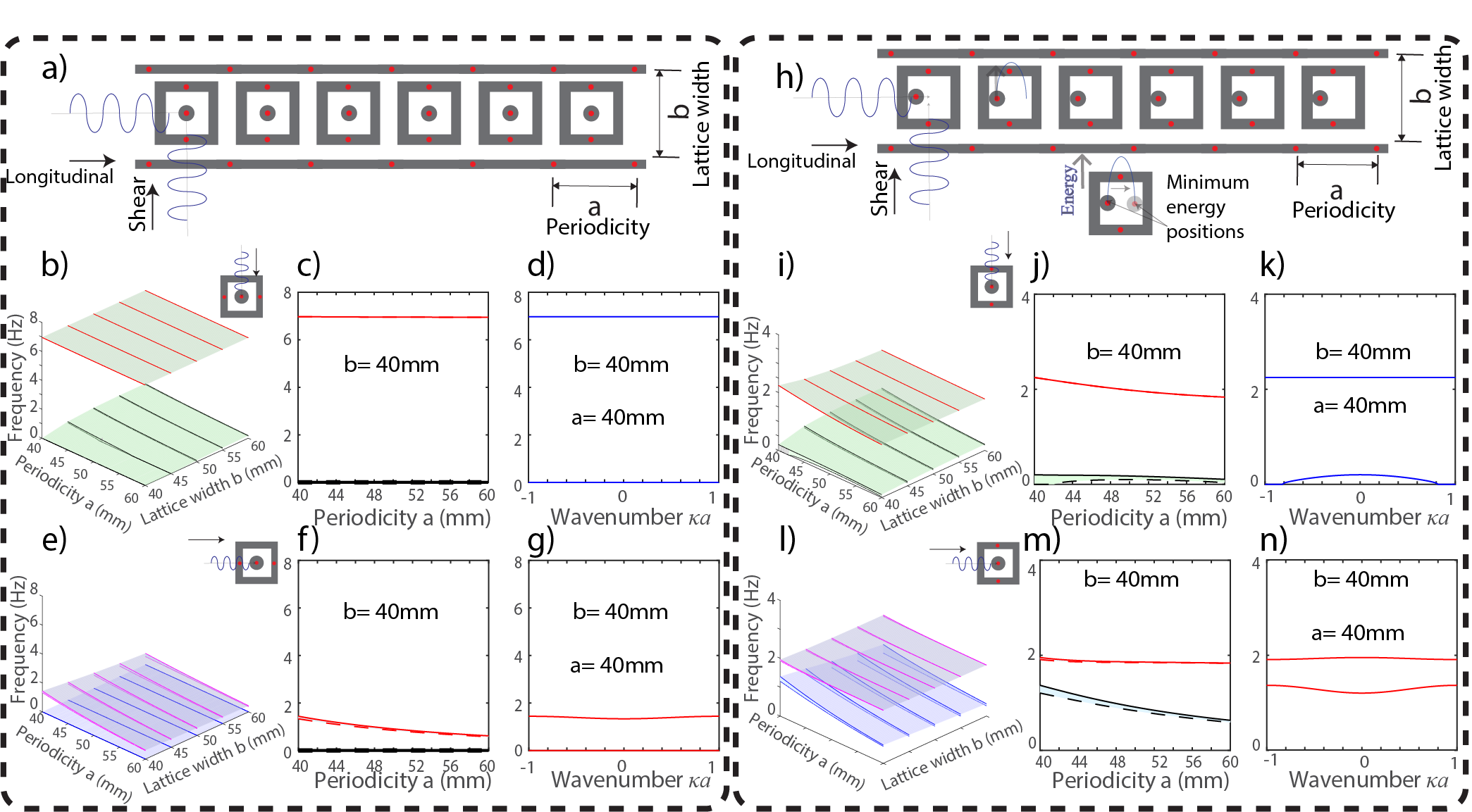} 
\caption{\label{Longitudinal wavenumber band gab}
\textbf{A study of the rectangular system's parameters in Design B} A schematic of the rectangular system with (a) the inner resonator at the center of the unit cell and (h) inner resonator at one of the two minimum energy positions. Three-dimensional parametric sweep of (b\& e) Design B-center resonator and (i\& l) Design B- sided resonator in both shear and longitudinal modes, respectively, showing the effect of varying both the lattice width and periodicity on the frequency range of transmission.  Dispersion frequency band width as a function of lattice periodicity with unit cell width = 40 $\mathrm{mm}$ for (c\& f) Design B-center resonator and (j\& m) Design B- sided resonator for shear and longitudinal modes, respectively.   
Dispersion relation (frequency vs. wavenumber) for unit cell width and length $ a = b = 40 \, \mathrm{mm} $  for (d\& g) Design B-center resonator and (k\& n) Design B- sided resonator for shear and longitudinal modes, respectively.}
\end{figure*}

In addition to opening a partial wavenumber band gap for the first shear branch, it is possible to engineer a complete wavenumber band gap for both longitudinal and shear modes. To achieve such feat, we consider Design B, where we embed two magnets at the top-center of the rectangular frame instead of four magnets at the corners in Design A. In Design B, there are two minimum energy resting positions for the circular disk (i.e., bistable unit cell design Fig. \ref{Longitudinal wavenumber band gab}h-inset). Such bistability can enable two possible dynamic behaviors or scenarios as a response to a harmonic excitation. Both of these possible scenarios are functions of the excitation amplitude and the strength of the energy barrier between the two stable potential wells. In case 1: the barrier between the two potential wells is shallow enough for the circular disk to oscillate back and forth between wells, crossing the rectangular frame magnets each time (Fig. \ref{Longitudinal wavenumber band gab}a-g). In case 2: the circular disk is confined to one of the potential energy wells and does not cross to the other side of the rectangular frame as it oscillates (Fig. \ref{Longitudinal wavenumber band gab}h-n). In the first case, the system has sufficient physical negative inter-stiffness between the moving frame and the inner disc, opening a full wavenumber band gap in the longitudinal direction  (Fig. \ref{Longitudinal wavenumber band gab}e-g). A full wavenumber band gap also emerges in the shear direction, due to the negative physical stiffness between the rectangular frames (Fig. \ref{Longitudinal wavenumber band gab}b-d). In the second case, where the disk does not cross between the different potential wells, only a partial wavenumber band gap can exist in the shear direction (Fig. \ref{Longitudinal wavenumber band gab}h-n).

To better understand the effect of hierarchy on our metamaterials, we explore the evolution of the dispersion relation as a function of the lattice parameters in the case of multiple magnets per unit cell substructure (i.e., Design C with four magnets per each concentric circle). In this case, the lattice periodicity dictates the stable position and orientation of each magnetic ring relative to the magnetic boundary and to each other. For a given lattice width, $b$ = 50 mm, the minimum energy position for the outer ring at lower periodicity (e. g., a = 40 mm) is at orientation angle of $45^\circ$ (Fig.~\ref{Parametric circular}c). In contrast, at higher periodicity values (e. g., a = 52 mm) the outer ring takes a stable orientation angle of $0^\circ$. In both cases, the inner ring self-orients to minimize its energy at $45^\circ$ offset from the outer one. The wavenumber band gap for Design B exists for the lowest shear mode at 49 $< a <$ 54 mm (Fig. \ref{Parametric circular}c). The higher shear branch is flat for most of the explored design search space. In addition, a zero frequency band gap exists for all periodicity values in the longitudinal direction and a full zero-frequency band gap for both modes for $a<48$ mm. 

It is worth noting that for Design C, we can obtain a two flat band for both shear and longitudinal modes at the same frequency within the same unit cell configuration. Such degeneracy plays an important role in localizing both modes simultaneously within the finite structure. Degenerate flat bands can be harnessed to localize orbital angular momentum (OAM) with multiple potential applications \cite{zhu2024flatbands}. Figure \ref{degenerate flat band}a presents the unit cell and its dispersion curves with two flat bands for the internal resonator at $f=11.7$Hz. A zoomed in view confirms that the frequency difference between the two branches is extremely small $\Delta f = 0.05$Hz.  The implications of such degeneracy can be seen in the time domain responses for different excitations (Fig. \ref{degenerate flat band}b–g). When the system is excited using longitudinal or shear polarization, the response of the first inner resonator exhibits strong spatial localization with no leakage to the bulk (Fig. \ref{degenerate flat band}b–e). When the system is excited with a mixed  mode (45$^\circ$), both wave polarizations are localized to the first unit cell (Fig. \ref{degenerate flat band}f-g). Despite the different excitation directions, the temporal and spatial responses remain localized.  The displacement trajectories for both the resonator and the main mass under mixed polarization are presented in Figure \ref{degenerate flat band}h-i. Instead of oscillating in a single direction, the first resonator and first mass follow an oval-like path in the $x-y$ plane, indicating orbital motion. For comparison, we also plot the displacement trajectories of the second unit cell (i.e., resonator and main mass) under mixed mode excitation (Fig. \ref{degenerate flat band}h-i). The amplitudes in the second unit cell are significantly smaller, confirming that the orbital angular momentum (OAM) response is primarily localized within the first unit cell. An additional pair of flat bands can be added to the dispersion curves by adding a disk with an embedded permanent magnet at its center (i.e., Design D) inside the inner ring (Fig. \ref{Parametric circular}i-n). Table \ref{tab:stiffness_comparison} lists a comprehensive comparison of the different stiffness for Designs A and C.  

\begin{figure*}[!h]
\centering\includegraphics[scale = 1]{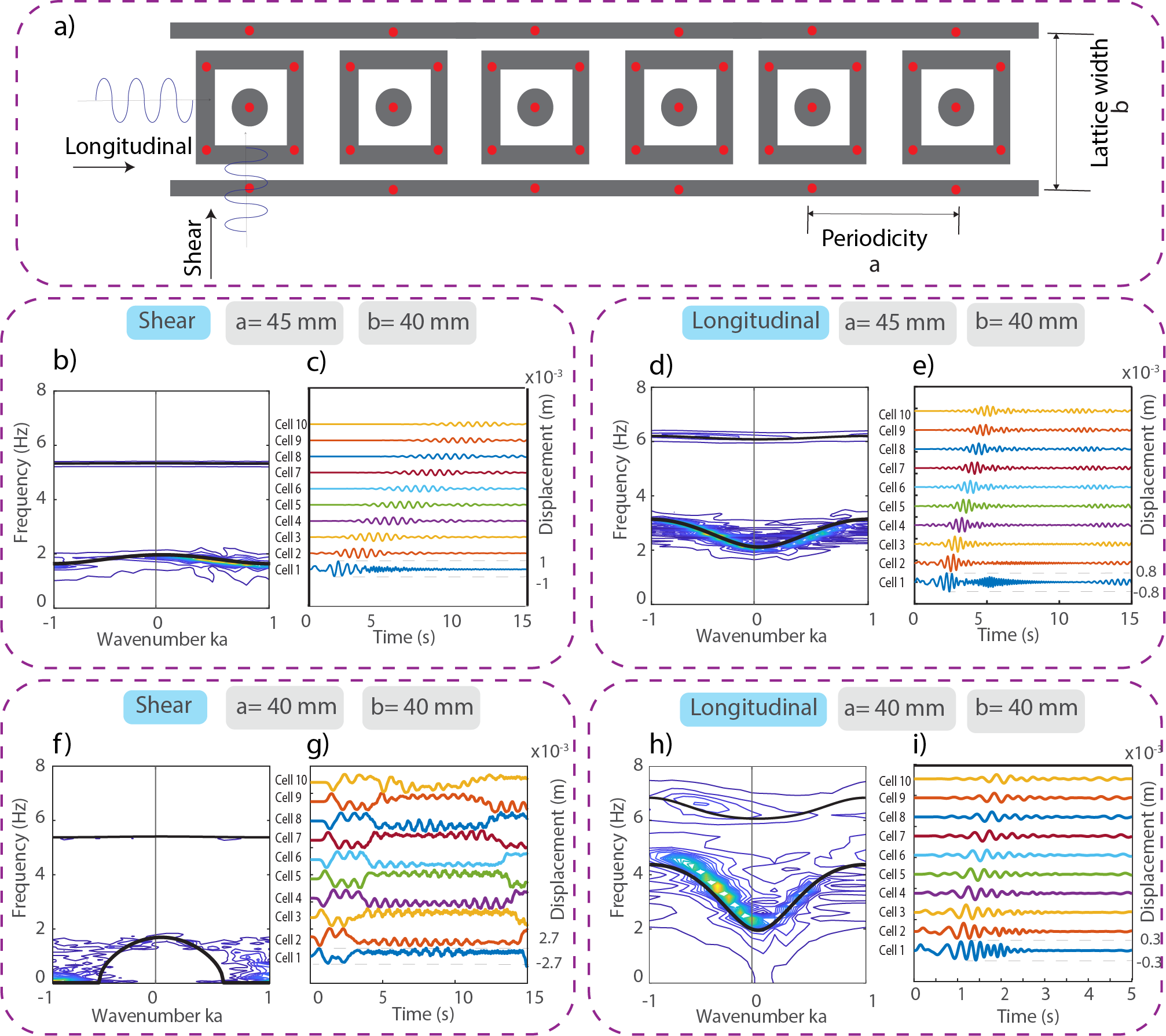}
\caption{\label{Nonlinear validation}\textbf{Numerical validation of analytical dispersion curves with flat bands, zero-frequency band gaps and wavenumber band gaps}. (a) A schematic of the rectangular system Design A with indicated shear and longitudinal direction excitations. Analytical shear dispersion curves (black lines) with superimposed 2D-FFT contours from the numerical excitation of the shear modes with  periodicity (b) $ a = 45 \, \mathrm{mm} $ and (f) $ a = 40 \, \mathrm{mm} $. The time -displacement history of the metamaterial frame and resonator with a periodicity (c) $ a = 45 \, \mathrm{mm} $ and (g) $ a = 40 \, \mathrm{mm} $. Analytical longitudinal dispersion curves (black lines) with superimposed 2D-FFT contours from the numerical excitation of the longitudinal modes with  periodicity (d) $ a = 45 \, \mathrm{mm} $ and (h) $ a = 40 \, \mathrm{mm} $. The time -displacement history of the metamaterial frame and resonator with a periodicity (e) $ a = 45 \, \mathrm{mm} $ and (i) $ a = 40 \, \mathrm{mm} $.  }
\end{figure*}

\begin{figure}[!h] 
\centering
\includegraphics[scale = 1]{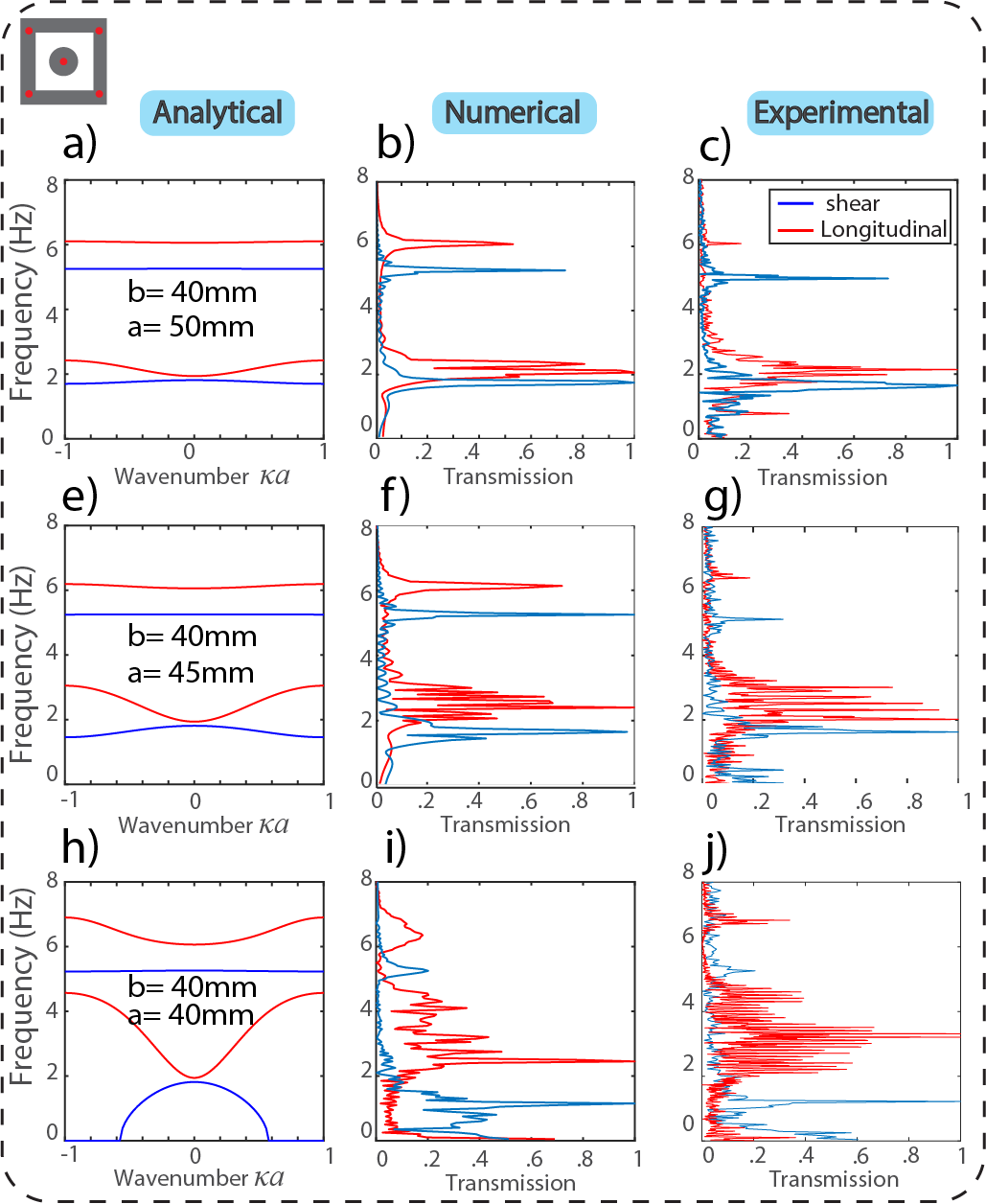}
\caption{\label{Experimental validation of the rectangular system}\textbf{Analytical, numerical, and experimental comparison of rectangular hierarchical system (Design A).} The infinite dispersion curves for unit cells with lattice width = 40 mm and periodicity (a) 50 mm, (e) 45 mm, and (h) 40 mm. The FFT of the numerically simulated displacement of the second frame (first resonator in case of localized wave) with lattice width = 40 mm and periodicity (b) 50 mm, (f) 45 mm, and (i) 40 mm. The FFT of the experimentally measured  displacement of the second frame (first resonator in case of localized wave) with lattice width = 40 mm and periodicity (c) 50 mm, (g) 45 mm, and (j) 40 mm. The blue lines correspond to the shear modes and the red lines correspond to longitudinal modes in all panels.}
\end{figure}

\begin{figure}[!b]
\centering
\includegraphics[scale = 1]{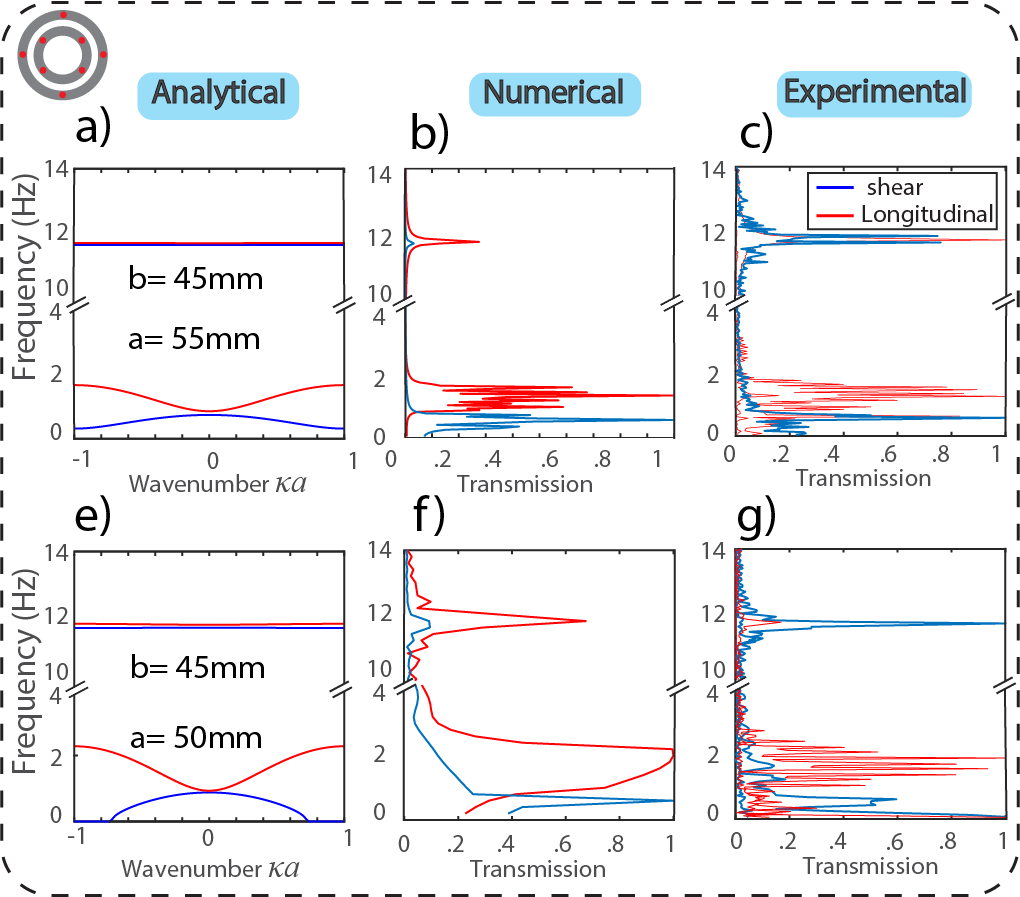}
\caption{\label{Experimental validation of the circular system}\textbf{Analytical, numerical, and experimental comparison of the circular hierarchical system (Design C).} The infinite dispersion curves for unit cells with lattice width = 45 mm and periodicity (a) 55 mm and (e) 45 mm. The FFT of the numerically simulated displacement of the second frame (first resonator in case of localized wave) with lattice width = 45 mm and periodicity (b) 55 mm and (f) 50 mm. The FFT of the experimentally measured  displacement of the tenth frame with lattice width = 45 mm and periodicity (c) 55 mm and (g) 55 mm. The blue lines correspond to the shear modes and the red lines correspond to longitudinal modes in all panels.}
\end{figure}

\begin{figure*}[!ht]   
\centering \includegraphics[scale = 1]{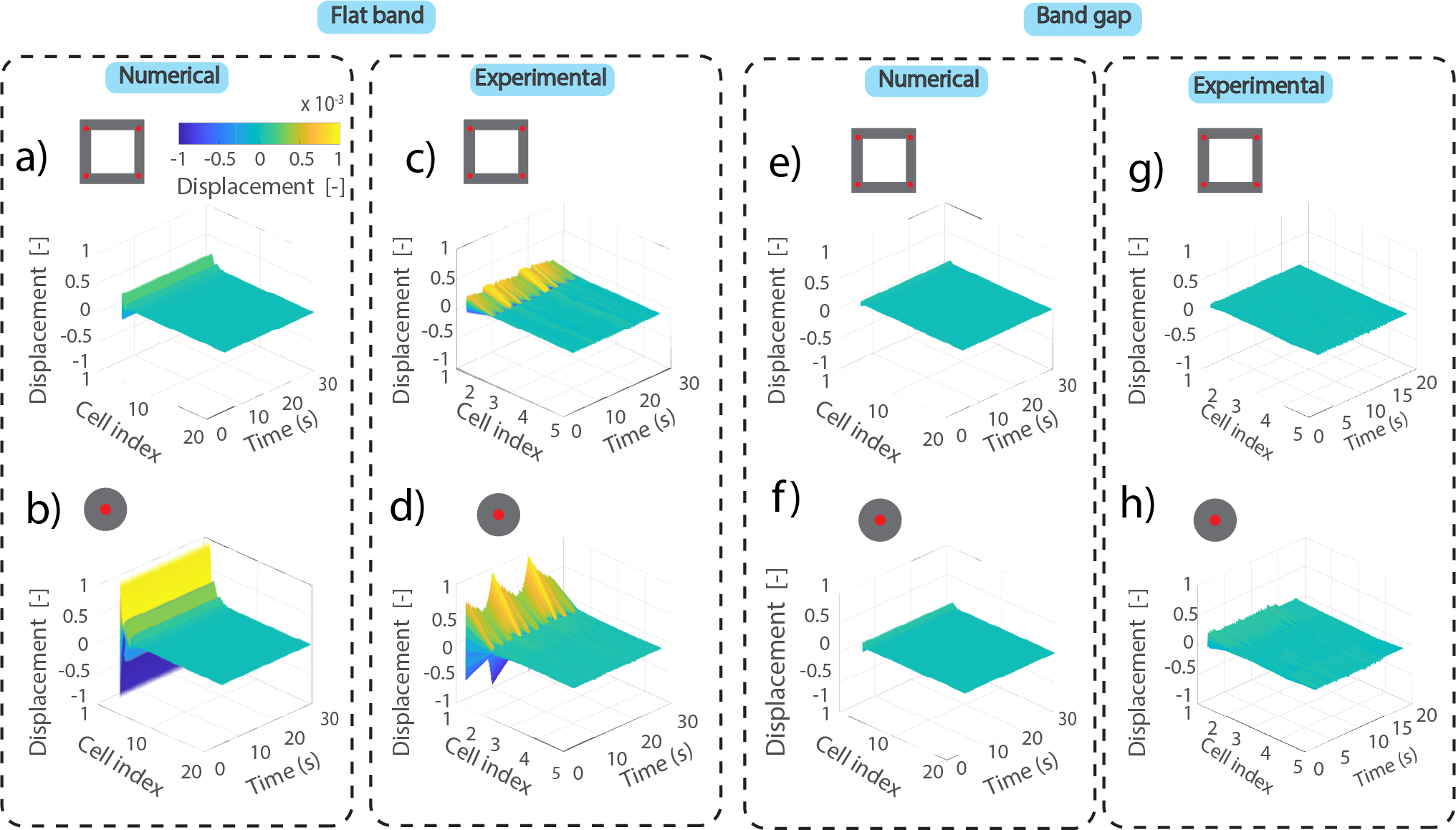}
\caption{\label{Flat band}\textbf{Numerical simulations and experimental validations of the rectangular system behavior at flat band vs band gap frequencies.} (a,b) Numerically simulated and (c,d) experimentally measured displacements over time of all (a,c) the outer-moving frames and (b,d) inner resonators as we excite the system at a flat band frequency of 6.193 Hz. 
(e,f) Numerically simulated and (g,h) experimentally measured displacements over time of all (e,g) the outer-moving frames and (f,h) inner resonators as we excite the system at a band gap frequency of 7 Hz.}
\end{figure*}

To validate our analytical infinite dispersion curves, we utilize numerical methods to simulate wave propagation within a finite metamaterial structure. We start by numerically solving the nonlinear equations (Eq. 2 and 3) for a finite structure composed of 20 unit cells. We excite the system at one end with a chirp signal between 0.01 and 10 Hz. We calculate the 2D-FFT of the displacement of all the frames and disks, separately, and overlay both contours on top of the analytically calculated dispersion curves. In addition to the dispersion curves, we plot the displacement of each disk as a function of time for the first 10 unit cells. We start with the simple case of $b=40$ mm and $a =45$mm. We observe excellent agreement between the finite and infinite models for both shear (Fig.~\ref{Nonlinear validation}b) and longitudinal excitations (Fig.~\ref{Nonlinear validation}d). The time signals in both cases show conventional oscillatory motion. For the case of $a = b = 40$ mm, where we predict a wavenumber band gap (Fig.~\ref{Nonlinear validation}f), we similarly see matching results. However, we observe a unique motion in the case of shear excitation due to the presence of the wavenumber band gap (Fig.~\ref{Nonlinear validation}g).     

To validate our analytical and numerical solutions, we experimentally characterize our finite metamaterials for both the circular and rectangular designs. We perform the experimental characterization for different lattice periodicities for each design to confirm the robustness and validity of our approach. We conduct the experiments on the surface of an air-bearing table to minimize friction and allow the metamaterial elements to interact based solely on their magnetic repulsion forces. Once we turn on the air bearing, all masses naturally move to their minimum energy positions ( Fig. \ref{fig:Concept}b-d). We utilize 3x3 mm permanent magnets in all designs (i.e., in the outer masses, inner resonators, and boundary magnets), with the exception of the inner disk in designs B and D, which has 3x1 mm magnet to maintain system stability. The magnetic force parameters in Eq. 4 are experimentally measured as  $A = 5.2202 \times 10^{-11}$  and  $\gamma = -4$  for the 3x3 magnets and  $A = 8.6978\times 10^{-12}$  and  $\gamma = -4.1973$  for the smaller 3x1 magnets. We excite the first unit cell using a chirp signal ranging from 0.01 Hz to 14 Hz, which covers the expected transmission region. We repeat each experiment two times, where we excite the metamaterial once in the $x$-direction (longitudinal) and another in the $y$-direction (shear). To capture the motion of the frames and disks, we utilize a bird’s-eye view camera positioned above the air-bearing surface. We employ the digital image correlation engine (DICe) to analyze the displacement of our metamaterial elements in time. For each unit cell, we compute the Fast Fourier Transform (FFT) of the displacement signals from all masses to extract the dynamic response. Figures \ref{Experimental validation of the rectangular system} and \ref{Experimental validation of the circular system}  provide a thorough comparison between analytical, numerical, and experimental approaches. In Figure \ref{Experimental validation of the rectangular system}(a-c), we excite a rectangular system with a periodicity $a = 50\,\mathrm{mm}$ in both directions, while in Figures \ref{Experimental validation of the rectangular system}(e-g and h-j), we change the periodicity of the system to $45\,\mathrm{mm}$ and $40\,\mathrm{mm}$, respectively, to validate the physics of the system in all possible scenarios and hence different dispersion cases. We repeat the previous steps for the circular system with two lattice constants $a = 55\,\mathrm{mm}$
and $a = 50\,\mathrm{mm}$ as shown in Figure \ref{Experimental validation of the circular system}(a-c and e-g), respectively. The results indicate a very good agreement between the analytical, numerical, and experimental approaches.

To further show the exciting unique features of our system, we demonstrate the attainable localization with a flat band frequency in contrast to a band gap frequency. In both cases, the waves do not propagate to the other end of the lattice, but the localization level is quite different. To better understand how the wave behaves at these frequencies, we excite the first unit cell at two different frequencies: a flat band at 6.19 Hz and a band gap at 7 Hz. We choose these two frequencies to be very close to eliminate any drastic frequency-dependent amplitude change. To help visualize the wave behavior within the lattice, we plot the displacement of all the lattice's outer (Fig. \ref{Flat band}a) and inner (Fig. \ref{Flat band}b) masses in the flat band case. The plots show extreme wave  localization at the first unit cell, with the inner mass showing a higher localization amplitude than the frame. In the case of band gap excitation, we see a much muted response, both for the inner (Fig. \ref{Flat band}e) and the outer masses (Fig. \ref{Flat band}f). The results from both simulations show the stark difference in behavior when we excite the system in a flat band vs. a band gap. To validate our numerical findings, we experimentally excite the same metamaterial at both frequencies and track the displacement of all of the masses using a bird-eye camera. The displacement profiles confirm (1) the extreme localization of the wave in the case of flat band (Fig. \ref{Flat band}c and d) compared to the band gap (Fig. \ref{Flat band}g and h) and (2) the higher level of localization in the inner mass (Fig. \ref{Flat band}d) compared to the outer mass (Fig. \ref{Flat band}c).

\section{\label{sect: Intro} CONCLUSIONS}
In this work, we demonstrate analytically, numerically, and experimentally the design and the dispersive properties of a passive tunable metamaterial hierarchical system. For the analytical solution, we use Bloch's wave solution to get the dispersion relations of the different systems. We also perform a parametric study on the system to determine the effect of varying the lattice parameters on the dispersive characteristics. The results indicate that reducing the periodicity of the system below a certain value can open a wavenumber band gap as a result of increasing the value of the negative physical stiffness between the cells. We also demonstrate that increasing the coupling stiffness between the main mass and the resonator within the same unit cell can flatten the dispersion curve resulting in a flat band. To validate the linear approximation, we utilize a nonlinear model to numerically solve for a finite number of unit cells. We obtain a good agreement between the two models. Finally, we experimentally validate our design approach, analytical predictions, and numerical simulations. We observe a good agreement between our experiments and both the numerical and analytical models. Our work paves the way for designing and utilizing exotic wave dispersion traits in advanced materials and devices. 

\bibliographystyle{unsrt}
\bibliography{references}
\newpage~
\newpage~

\beginsupplement

\begin{table*}
\centering
\setlength{\arrayrulewidth}{1pt}
\caption{\textbf{Comparison of longitudinal and shear stiffness components for the Rectangular system and the Circular system at unit width \textbf{$ b = 50 \, \mathrm{mm} $  }, and periodicity \textbf{$ a = 50 \, \mathrm{mm} $  } and \textbf{$ a = 55 \, \mathrm{mm} $}.}}

\begin{tabular}{|m{12mm}|c|c|c|c|c|c|c|}
\hline
\textbf{System} & \textbf{$a$} & $k_{\text{11}}^{(L)}$ & $k_{\text{12}}^{(L)}$ & $k_{\text{g1}}^{(L)}$ & $k_{\text{11}}^{(S)}$ & $k_{\text{12}}^{\text{(S)}}$ & $k_{\text{g1}}^{\text{(S)}}$ \\
\hline
Rect. & 50 & 0.1186 & 0.4499 & .0839 & -0.0191 & 0.2089 & 0.1698 \\
                     & 55 & 0.0538 & 0.4499 & 0. 0839& -0.0088 & 0.2089 & 0.1698 \\
\hline
Cir. & 50 & 0.1566 & 4.8215 & 0.0536 & -0.0324 & 4.8215 & 0.0536 \\
              & 55 & 0.0636 & 4.8215 & 0.0394 & -0.0142 & 4.82 & 0.0384 \\
\hline
\end{tabular}
\label{tab:stiffness_comparison}
\end{table*}

\begin{figure*}   \centering \includegraphics[scale = .95]{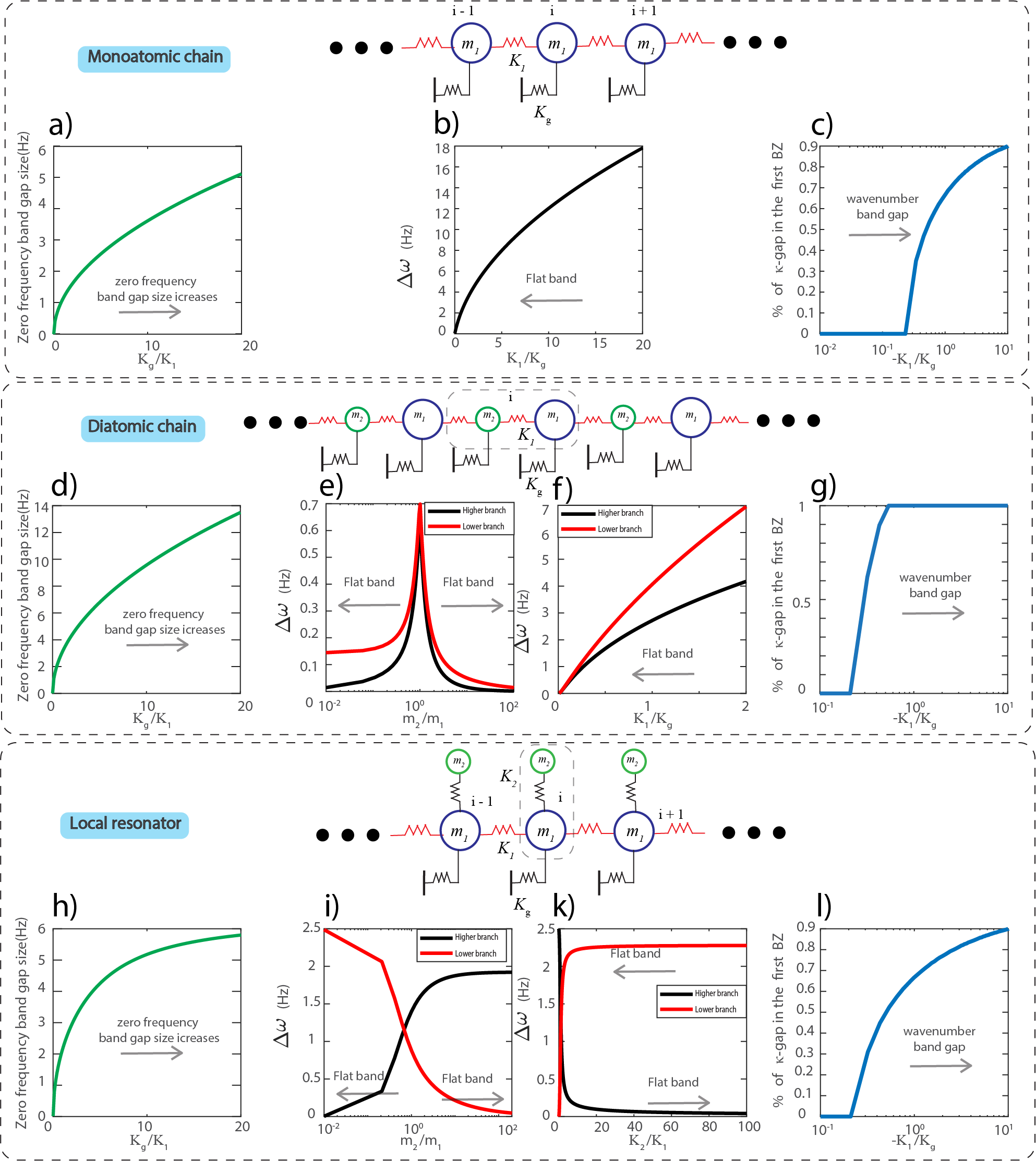}
\caption{\label{methodology}\textbf{Design methodology for opening a zero frequency band gap, getting a flat band, and opening a wavenumber band gap} (a) The relation between the ground stiffness and the zero frequency band gap size for a monoatomic chain. (b) The effect of decreasing of the inter stiffness between the masses of the monoatomic chain on the band width in order to get a flat band.  (c) The relation between the value of the physical negative stiffness to the ground stiffness and the closed wavenumber range for a monoatomic chain.(d) The relation between the ground stiffness and the zero frequency band gap size for a diatomic chain. (e) The effect of changing the impedance between the masses in a diatomic chain on the width of the higher and the lower dispersion curves. (f) the effect of Changing the inter stiffness between the masses and the width of the higher and the lower dispersion branches. (g) The relation between the value of the physical negative stiffness to the ground stiffness and the closed wavenumber range for a diatomic chain. (h) The The relation between the ground stiffness and the zero frequency band gap size for a local resonance system. (i) The effect of changing the impedance between the main and the resonator masses in a local resonance system on the width of the higher and the lower dispersion curves. (k) The relation between the inter stiffness $k_2$  and the width of the higher and the lower dispersion branches. (l) The relation between the value of the physical negative stiffness to the ground stiffness and the closed wavenumber range.   }
\end{figure*}

\vspace{-13mm}
{\large{Supporting information}}\\
In order to elucidate the methodology of tailoring the dispersion curves of our metamaterials (Fig. \ref{methodology}), we first establish the governing equations and the corresponding dispersion relations for spring-mass system representing  (1) monoatomic, (2) diatomic, and (3) locally resonant lattices. Then, we explore the effect of changing the lattice parameters on the desired dispersion traits such as flat dispersion bands, wavenumber band gaps, and zero-frequency band gaps. 

For a monoatomic chain, we can obtain a zero-frequency band gap, flat band, and a wavenumber band gap following simple strategies. We can open a zero-frequency band gap by introducing a ground stiffness to the system (Fig.\ref{methodology}a). We can increase the size of the zero-frequency band gap by increasing the value of the ground stiffness. To get a flat band we can reduce the coupling stiffness between the unit cells to further localize the wave (Fig. \ref{methodology}b). We can open a wavenumber band gap by introducing sufficient negative physical stiffness to the system (Fig.\ref{methodology}c). The width of wavenumber band gap increases as we increase the negative stiffness ratio until we close all the wavenumbers (i.e., a full wavenumber band gap). 
 
We can follow the same design principle for a diatomic chain. To open a zero frequency band gap, we introduce ground stiffness to the system (Fig. \ref{methodology}d). To get a flat band in a diatomic chain, we can introduce a higher impedance mismatch between the two masses inside the unit cell. The higher the difference in impedance,  the flatter the band becomes (Fig. \ref{methodology}e). We can also flatten the dispersion curves by reducing the inter-stiffness between masses in the unit cells (Fig. \ref{methodology}f). 

For a local resonant system, we can follow the ground stiffness strategy for opening a zero-frequency band gap (Fig.\ref{methodology}h). We can flatten the dispersion curve of the resonator (i.e., the higher dispersion branch) by reducing the mass ratio of the system  to introduce a difference in the impedance between the mass and its resonator (Fig. \ref{methodology}i). Another effective method is increasing the connecting stiffness between the main mass and the resonator (Fig. \ref{methodology}k). For the wavenumber band gap, we can use sufficient negative inter stiffness (Fig.\ref{methodology}l). 

\subsection*{A. Monoatomic chain}
We consider a periodic monoatomic lattice with mass $m_1$, inter-stiffness $K_{\mathrm{1}}$, lattice constant $a$, and ground stiffness $K_g$. The equation of motion for the $n^{th}$ mass is:
\begin{align}
m_1\ddot{x}_n + K_{\mathrm{1}}(2x_n - x_{n+1} - x_{n-1}) + K_g x_n = 0 .
\end{align}
Assuming a Bloch's solution in the form $x_n = U e^{i(\omega t - \kappa n a)}$, where $\kappa$ is the wavenumber, $U$ is the displacement, $\omega$ is the frequency, and $t$ is time. Substituting into $(\mathrm{S1})$ gives the dispersion relation:
\begin{align}
(-m_1\omega^2  + K_{\mathrm{1}}(2 - e^{i \kappa a} - e^{-i \kappa a}) + K_g) U = 0.
\end{align}
In order to get the ratio between the negative inter-stiffness and the ground stiffness at which we can open a wavenumber band gap, we derive the characteristic equation. The intersection between the dispersion curve and the wavenumber axis for the first Brillouin zone with the presence of ground stiffness, represents the wavenumber band gap range (i.e. at $\omega = 0$)
\begin{align}
\kappa_{\omega=0} = \frac{1}{a} cos^{-1} (1+\frac{K_{g}}{2K_{1}})
\end{align}
The previous equation (S3) has a mathematical constraint:
\begin{align}
-1<(1+\frac{K_{g}}{2K_{1}})<1
\end{align}
We can simplify the previous constraint to get the sufficient stiffness ratio at which we can open a wavenumber band gap: 
\begin{align}
-2<\frac{K_{g}}{2K_{1}}&<0\\
-\frac{1}{4}>\frac{K_{1}}{K_{g}}
\end{align}
\subsection*{B. Diatomic chain}
For a diatomic chain with masses $m_1,m_2$, equal inter-cell stiffness $K_{\mathrm{1}}$, and ground stiffness $K_g$, the equations of motion for the $n$$^{th}$ unit cell (displacements $x^n_1,x^n_2$) are
\begin{align}
m_1\ddot{x}^n_1 + K_{\mathrm{1}}(2x^n_1 - x^n_2 - x^{n-1}_2) + K_g x^n_1 &= 0, \\
m_2\ddot{x}^n_2 + K_{\mathrm{1}}(2x^n_2 - x^n_1 - x^{n+1}_1) + K_g x^n_2 &= 0. 
\end{align}
With the Bloch’s solution
\[
x^{n}_1 = u_1 e^{i(\omega t - \kappa n a)},\qquad
x^{n}_2 = u_2 e^{i(\omega t - \kappa n a)},
\]
the equations become:
\begin{align}
-m_1\omega^2 u_1 + K_{\mathrm{1}}(2u_1 - u_2 - u_2 e^{-i k a}) + K_g u_1 &= 0, \\
-m_2\omega^2 u_2 + K_{\mathrm{1}}(2u_2 - u_1 - u_1 e^{+i k a}) + K_g u_2 &= 0. 
\end{align}
Rearranging into matrix form yields:
\begin{equation}
\begin{bmatrix}
2K_{\mathrm{1}} + K_g - m_1\omega^2 & -K_{\mathrm{1}}(1 + e^{-i k a}) \\
- K_{\mathrm{1}}(1 + e^{+i k a}) & 2K_{\mathrm{1}} + K_g - m_2\omega^2
\end{bmatrix}
\begin{bmatrix} u_1 \\[6pt] u_2 \end{bmatrix}
=
\begin{bmatrix} 0 \\[6pt] 0 \end{bmatrix}.
\end{equation}
Where we can get the dispersion relation for the system by getting the determinate of the previous matrix. By setting $\omega = 0$ in the characteristic equation, we can get the intersection between the dispersion curve and the wavenumber axis for the first Brillouin zone:
\begin{align}
\kappa_{\omega=0} = \frac{1}{a} cos^{-1}\big(\frac{1}{2} \big((\frac{K_{g}}{K_{1}})^{2}+\frac{4K_{g}}{K_{1}}+2\big)\big)
\end{align}

Again, by applying the mathematical constraint, we can get the stiffness ratio as:
\begin{align}
-1<&\frac{1}{2} \big((\frac{K_{g}}{K_{1}})^{2}+\frac{4K_{g}}{K_{1}}+2\big)<1\\
-\frac{1}{4}>&\frac{K_{1}}{K_{g}}
\end{align}

\subsection*{C. Locally resonant chain}
We also consider a chain where each primary mass $m_1$ is coupled to a local resonator of mass $m_2$ by stiffness $K_2$. Inter-cell coupling is $K_{\mathrm{1}}$ and ground stiffness is $K_g$. Let $u^n$ denote the main mass displacement and $v^n$ the local resonator displacement in cell $n$. The equations of motion are:
\begin{align}
m_1 \ddot{u}^n 
&+ K_{\mathrm{1}}(2u^n - u^{n+1} - u^{n-1})
+ K_g u^n \nonumber\\
&+ K_2(u^n - v^n) = 0, \\
m_2 \ddot{v}^n 
&+ K_2(v^n - u^n) = 0. 
\end{align}
We then apply the Bloch form of the solutions $u^n = U e^{i(\omega t - n \kappa a)},\; v^n = V e^{i(\omega t - n \kappa a)}$. By Substituting, we get: 
\begin{align}
\big(-m_1\omega^2 
&+ 2K_{\mathrm{1}}(1-\cos \kappa a) 
+ K_g 
+ K_2\big)U -
K_2 V &= 0, \\
- K_2 U +
&(-m_2\omega^2 + K_2)V &= 0. 
\end{align}
Or in matrix form:
\begin{equation}
\begin{aligned}
\begin{bmatrix}
2K_{\mathrm{1}}(1-\cos \kappa a) + K_g + K_2 - m_1\omega^2 & -K_2 \\
- K_2 & K_2 - m_2\omega^2
\end{bmatrix}
\begin{bmatrix} U \\ V \end{bmatrix}
\\
=~
\begin{bmatrix} 0 \\ 0 \end{bmatrix}.
\end{aligned}
\end{equation}

Again, the dispersion relation is obtained from the determinate of the previous matrix. By setting $\omega = 0$ in the characteristic equation, we can get the point of the dispersion curve on the wavenumber axis  for the local resonant system as following:
\begin{align}\tag{S20}
\kappa_{\omega = 0} = \frac{1}{a} cos^{-1} (1+\frac{K_{g}}{2K_{1}})
\end{align}
Thus the sufficient stiffness ratio at which we can open a wavenumber band gap is:
\begin{align}
-2<\frac{K_{g}}{2K_{1}}&<0\\
-\frac{1}{4}>\frac{K_{1}}{K_{g}}
\end{align}

\end{document}